\documentstyle[psbox]{WORKSHOP}
\begin{document}

\title{Groups of Galaxies at Intermediate Redshift}

\author{
Eric D.~Miller,$^1$
Marshall Bautz,$^1$
Catherine Grant,$^1$ \\
Ryan Hickox,$^2$
Mark Brodwin,$^2$
Stephen Murray,$^2$
Christine Jones,$^2$ \\
William Forman$^2$ \&
Alexey Vikhlinin$^2$ 
\\[12pt]  
%
$^1$  MIT Kavli Institute for Astrophysics and Space Research, Cambridge,
MA, USA \\
$^2$  Harvard-Smithsonian Center for Astrophysics, Cambridge, MA, USA \\
%
{\it E-mail(EDM): milleric@mit.edu} 
}

\abst{
Galaxy groups are key tracers of galaxy evolution, cluster evolution, and
structure formation, yet they are difficult to study at even moderate
redshift.  We have undertaken a project to observe a flux-limited sample of
intermediate-redshift ($0.1 < z < 0.5$) group candidates identified by the
XBootes Chandra survey.  When complete, this project will nearly triple the
current number of groups with measured temperatures in this redshift range.
Here we present deep Suzaku/XIS and Chandra/ACIS follow-up observations of
the first 10 targets in this project; all are confirmed sources of
diffuse, thermal emission with derived temperatures and luminosities
indicative of rich groups/poor clusters.  By exploiting the
multi-wavelength coverage of the XBootes/NOAO Deep Wide-Field Survey
(NDWFS) field, we aim to (1) constrain non-gravitational effects that
alter the energetics of the intragroup medium, and (2) understand the
physical connection between the X-ray and optical properties of groups. We
discuss the properties of the current group sample in the context of
observed cluster scaling relations and group and cluster evolution and
outline the future plans for this project.
}

\kword{galaxies: clusters --- X-rays: galaxies: clusters --- surveys}

\maketitle
\thispagestyle{empty}

\section{Motivation}

Galaxy groups are vital to our understanding of structure formation,
cluster evolution, and galaxy evolution.  
In the local universe, 50\%--70\% of all galaxies are found in groups
(Tully 1987).  
Under the standard picture of hierarchical structure formation, groups merge
to form clusters.  Since group masses are relatively low, they are ideal
sites in which to observe non-gravitational processes that affect the
energetics of the plasma in both groups and clusters (e.g.~Balogh 2006).
Interactions among group galaxies, as well as interactions
between individual galaxies and the group gravitational potential and
intragroup medium (IGM), can alter galaxy properties substantially
(Mulchaey 2000, Rasmussen et al.~2006).  

Groups are difficult to study at even moderate redshifts ($z > 0.1$)
because the galaxy overdensity is low and because X-ray luminosities are
modest ($L_{X} \sim 10^{41}$--$10^{43}$ erg s$^{-1}$).  The XBootes 
Chandra survey (Murray et al.~2005) provides a powerful opportunity for
systematic study of more distant groups.  The 9.3 deg$^{2}$\ survey region
is almost fully covered by deep optical and near-IR imaging from the NOAO
Deep Wide-Field survey (NDWFS; Jannuzi \& Dey 1999), by the Spitzer/IRAC
Shallow Survey (Eisenhardt et al.~2004), and by optical spectroscopy of
over 20,000 galaxies from the AGN and Galaxy Evolution Survey (AGES) and
ongoing MMT observations.

We have undertaken a project to observe a flux-limited sample of
intermediate-redshift ($0.1 < z < 0.5$) group candidates identified by the
XBootes Chandra survey (Kenter et al.~2005). By exploiting the
multi-wavelength coverage of the XBootes/NDWFS field, we aim to constrain
non-gravitational effects that alter the energetics of the IGM, and to
understand the connection between the X-ray and optical properties of
groups.  Here we present deep Suzaku/XIS and Chandra/ACIS follow-up
observations of the first targets in this project.

\section{Sample Selection}

\begin{figure}[ht]
\centering
\psbox[xsize=.75\linewidth]{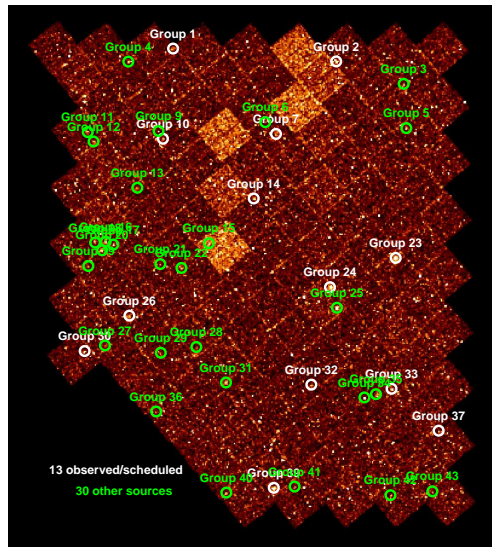}
\caption{Counts image mosaic of the original 5 ksec XBootes Chandra
snapshot (Kenter et al.~2005).  The 13 confirmed groups in our current
sample are identified in white, the remaining 30 extended X-ray sources are
circled in green.  The image has been smoothed, so some bright point sources
appear extended.  A handful of fields suffered from high background,
leading to the patchy appearance.}
\label{fig:xbootes}
\end{figure}

Of the 43 extended X-ray sources identified by the XBootes survey 
(see Figure \ref{fig:xbootes}), 27 exceed our flux threshold of
$2\times10^{-14}$ erg s$^{-1}$ cm$^{-2}$.  For 17 of these sources, we have
confirmed the presence of a bound group of galaxies with MMT spectroscopy.
The brightest 13 targets of the total sample have been observed (10 groups)
or awarded time (3 groups) with either Suzaku or Chandra, depending on the
presence of contaminating X-ray point sources.  These targets are listed
in Table \ref{tab:targets}.  We continue to obtain redshift information for
the remaining sample, and additional groups will be proposed for
observation in future cycles of Suzaku and Chandra.

\begin{table*}[t]
\caption{Group target list and observations \label{tab:targets}}
\begin{center}
\begin{tabular}{cccrccc} 
\hline\hline\\[-6pt]
Group        & XBootes ID             & $z$   & $S_{14}^a$ & Telescope     & Date Obs.     & ksec \\
\hline
\phantom{0}1 & CXOXB J143449.0+354301 & 0.151 &  4.2     & Suzaku  &  Dec 2007     & 42 \\
\phantom{0}7 & CXOXB J143109.1+350609 & 0.194 &  4.7     & Suzaku  &  June 2007     & 42 \\
30           & CXOXB J143747.6+333110 & 0.222 &  4.1     & Suzaku  &  June 2007     & 39 \\
\phantom{0}2 & CXOXB J142900.6+353734 & 0.234 &  6.9     & Chandra &  May 2007     & 38 \\
26           & CXOXB J143615.4+334650 & 0.342 &  4.5     & Suzaku  &  June 2007     & 44 \\
14           & CXOXB J143156.1+343806 & 0.350 &  4.2     & Chandra &  Sept 2008     & 50 \\
24           & CXOXB J142916.1+335929 & 0.131 &  9.7     & Chandra &  Nov 2008     & 25 \\
10           & CXOXB J143508.8+350349 & 0.280 &  3.3     & Chandra &  Nov 2008     & 45 \\
37           & CXOXB J142532.9+325644 & 0.215 &  2.3     & Suzaku  &  July 2008     & 40 \\
32           & CXOXB J142955.8+331711 & 0.420 &  2.7     & Suzaku  &  Feb 2009     & 40 \\
\hline
33           & CXOXB J142709.3+331510 & 0.242 & 6.3     & Chandra &  AO10 (awarded)          & 30 \\
23           & CXOXB J142657.9+341201 & 0.130 & 22.1     & Chandra &  AO10 (awarded)          & 20 \\
39           & CXOXB J143113.8+323225 & 0.278 & 11.2     & Chandra &  AO10 (awarded)          & 50 \\
\hline
\end{tabular}
\\
$^a$ XBootes survey flux (0.5--2 keV) in units of 10$^{-14}$ erg s$^{-1}$
cm$^{-2}$
\end{center}
\end{table*}

Due to the size and depth of the XBootes survey, this sample of groups
fills a niche not covered by other surveys.  In general, our groups are
poorer/cooler than the clusters found in the larger but shallower ROSAT
400d survey (Burenin et al.~2007).  They should be richer than those
detected with XMM in the COSMOS survey (Finoguenov et al.~2007).  When
complete, this project will triple the current number of
intermediate-redshift groups with measured temperatures.

\section{Observations and Spectral Modeling}

\begin{figure*}[t]
\centering
\begin{minipage}[l]{.45\linewidth}
\centering
\psbox[xsize=\linewidth]{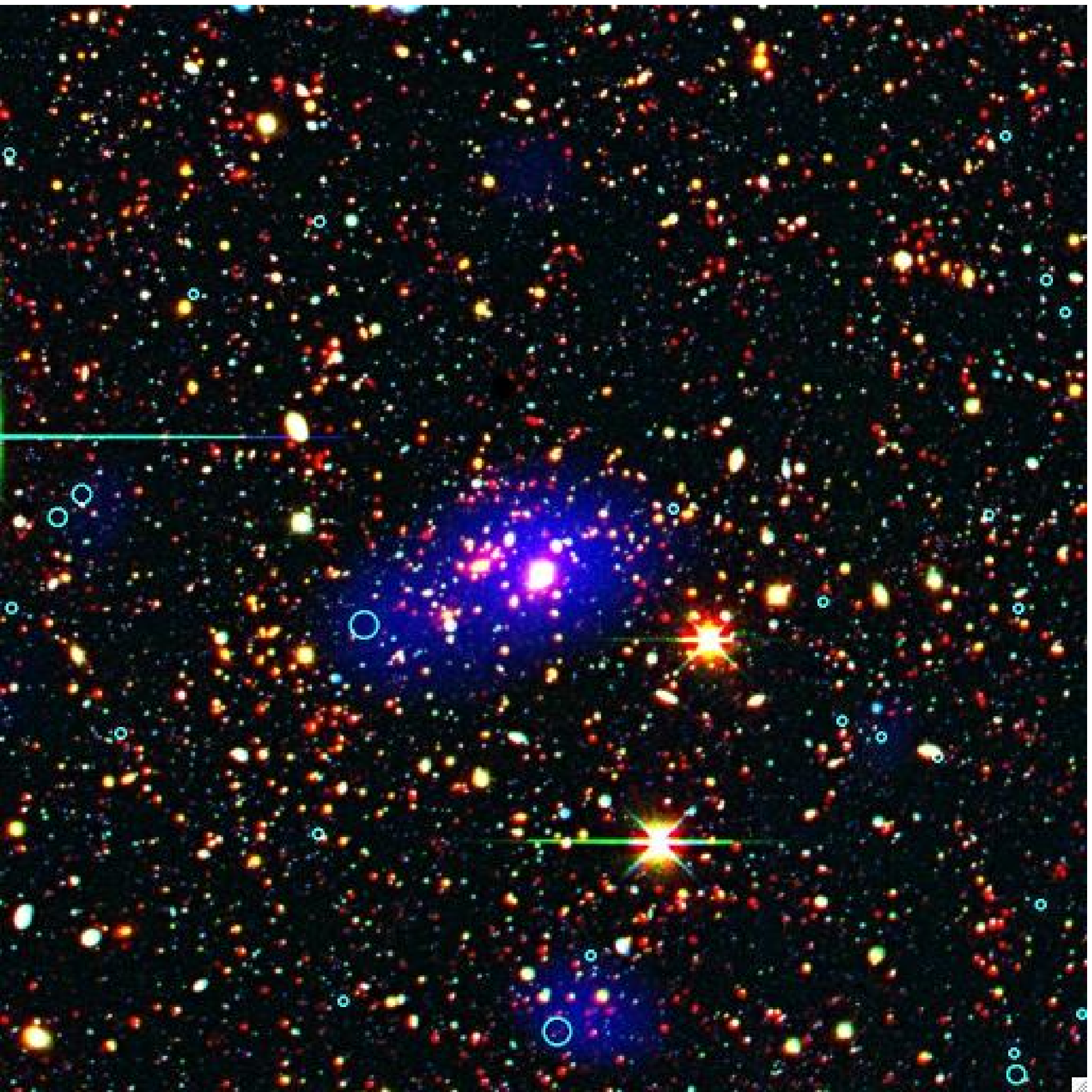}
\caption{Image of Group 1 ($z = 0.151$, $kT = 0.67$ keV) showing the NDWFS
$B_W$,$I$ and Spitzer/IRAC 3.6 $\mu$m combined image.  The Suzaku/XIS X-ray
emission is shown as diffuse blue color.  XBootes point sources are
identified by open circles; the diffuse group X-ray emission appears
extended to the east due to a nearby point source.  North is up, and the
image is 10 arcmin $\sim$ 1.6 Mpc on a side.}
\label{fig:im1}
\end{minipage}
\begin{minipage}[c]{.10\linewidth}
\end{minipage}
\begin{minipage}[r]{.45\linewidth}
\centering
\psbox[xsize=\linewidth]{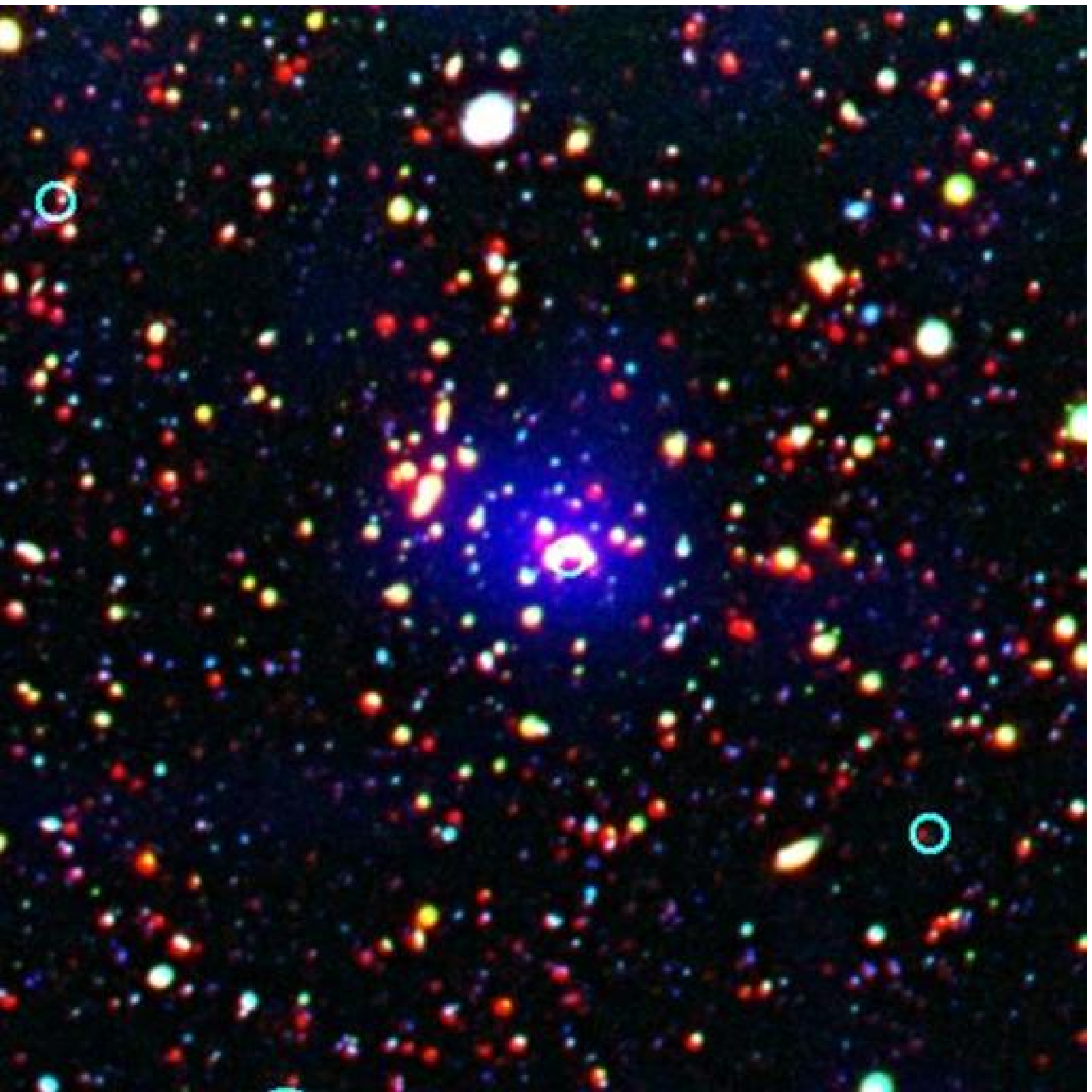}
\caption{Image of Group 14 ($z = 0.352$, $kT = 1.45$ keV) showing the NDWFS
$B_W$,$I$ and Spitzer/IRAC 3.6 $\mu$m combined image.  The Chandra/ACIS X-ray
emission is shown as diffuse blue color.  XBootes point sources are
identified by open circles; all X-ray point sources have been excluded.
North is up, and the image is 5 arcmin $\sim$ 1.5 Mpc on a side.
\vspace{\baselineskip}
}
\label{fig:im2}
\end{minipage}
\end{figure*}

We detect extended soft X-ray emission from all 10 observed targets, with 
0.5--2 keV fluxes ranging from 3--21 $\times 10^{-14}$ erg s$^{-1}$
cm$^{-2}$.  For some sources this is nearly five times the measured XBootes
snapshot flux (given in Table \ref{tab:targets}), although this is expected
from the deeper exposures.  Sample images from Suzaku and Chandra are shown
in Figures \ref{fig:im1} and \ref{fig:im2}, with the exposure-corrected
X-ray image overlaid on a combined NDWFS $B_W$,$I$ and Spitzer/IRAC 3.6
$\mu$m image.  The Suzaku image in Figure \ref{fig:im1} contains X-ray
point sources; the Chandra image in Figure \ref{fig:im2} has been smoothed
after point source removal.

The Suzaku/XIS group spectra were extracted from circular apertures of 1
Mpc radius at the group redshift. Point sources brighter than 10$^{-14}$
erg s$^{-1}$ cm$^{-2}$ were masked out. Due to vignetting and non-uniform
OBF contamination, the X-ray background was fit simultaneously to a region
outside the group aperture. 
Detector background was corrected using the accumulated Suzaku night
Earth background data.  The Chandra/ACIS spectra were extracted from a
similar aperture, with a nearby region used as the background.
Point sources were masked by hand.  

The diffuse group emission was modeled using the APEC plasma code
with variable temperature and abundance.  For all groups, the best-fit $kT$
ranges from 0.7--2.5 keV with abundances of 0.1--0.7 solar.  The range of
temperatures is illustrated in Figure \ref{fig:lxtx}.

\section{Initial Results: The $L_X$-$T_X$ Relation}

\begin{figure*}[t]
\centering
\psbox[xsize=.8\linewidth,rotate=r]{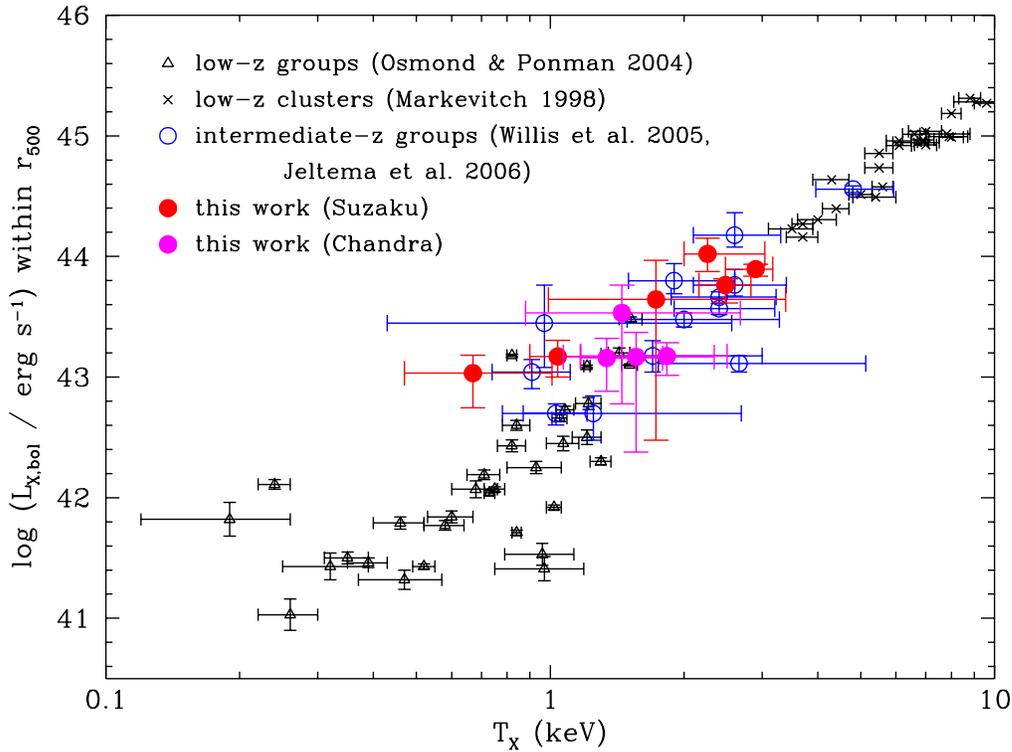}
\caption{$L_X$-$T_X$ relation for our 10 groups, plotted with low-$z$ group
and cluster samples and a sample of 11 other intermediate-$z$ groups with
measured $T_X$.  The XBootes systems lie in a region of rich groups and poor
clusters.  No aperture correction has been performed on the XBootes group
emission, therefore an additional systematic error of $<$ 50\% is
possible in $L_{X,{\rm bol}}$.}
\label{fig:lxtx}
\end{figure*}

Scaling relations identify divergence from self-similarity due to
non-gravitational effects (pre-collapse heating, galactic/AGN feedback,
radiative cooling).  Evolution in scaling relations at group (rather than
cluster) scales is a powerful diagnostic because these non-gravitational
effects are more important at smaller mass scales.  A small number of
groups have been observed at intermediate redshift with XMM-Newton (Willis
et al.~2005, Jeltema et al.~2006), and they show little if any evolution in
the $L_X$-$T_X$ scaling relation (see Figure \ref{fig:lxtx}).  A number of
other groups at $z > 0.15$ have measured $T_X$ (Bauer et al.~2002, Grant et
al.~2004, Fassnacht et al.~2007), bringing the total to 16, excluding this
work.  The large intrinsic scatter in the $L_X$-$T_X$ relation requires a
large sample of groups to distinguish between various models. Our full
sample will triple the number of groups with measured $T_X$ in this
redshift range. The first 10 groups have properties consistent with the
observed $L_X$-$T_X$ relation at $z \sim 0$ (see Figure \ref{fig:lxtx}).
They lie in a region on the faint end of the cluster population and bright
end of the typical group population, similar to the XMM-Newton group
samples.

The XBootes group $L_X$-$T_X$ relation shows a hint of flattening at the
low $kT$ end, however systematic effects prevent us from concluding this at
the present.  No spatial analysis has been completed to extrapolate the
group X-ray fluxes out to $r_{500}$.  We expect the aperture correction to
be less than 50\%, and this will increase the $L_X$ systematically.  On the
other hand, the Suzaku data suffer from AGN contamination due to the poor
spatial resolution.  This correction will decrease $L_X$ for the six groups
observed with Suzaku.

\section{Future Work: Evolution of the Group X-Ray/Optical Connection}

Most studies of the properties of groups beyond the local universe rely
on X-ray selected samples, since finding the extended X-ray emission
indicative of a virialized system is not difficult in surveys with
sufficient depth.  
On cluster scales, it is clear that there are
systematic differences in the X-ray properties depending on sample
selection method (e.g.~Donahue et al.~2001, Lubin et al.~2004, Barkhouse et
al.~2006).  On group scales ($kT < 3$ keV), only half of optically-selected
groups at low redshift are seen to produce X-ray emission (Mulchaey 2000,
Rasmussen et al.~2006).  This could be due to incomplete gravitational
collapse, a shallower potential (and lower X-ray temperature) than
expected, or simply a lack of IGM gas, perhaps as a result of feedback from
group galaxies to the IGM (Rasmussen et al.~2006).  
These differences likely result from the details of group formation
and the interplay between galaxies and the IGM, but they have been very
difficult to understand because we lack appropriate X-ray and optical data
on representative group samples beyond the local universe.

We have begun an optical search for groups using AGES and additional
MMT spectroscopic galaxy redshifts.  With follow-up data of the full X-ray
sample, we will directly compare the properties of optically-selected and
X-ray-selected groups, and we will be able to directly compare groups at
intermediate redshift to those at $z \sim 0$.  This will enhance larger
surveys such as the 400d cluster survey (Burenin et al.~2007), which 
is not sensitive to groups of this mass in this redshift range.  It will
also complement the deeper COSMOS survey (Finoguenov et al.~2007), which is
sensitive to lower mass groups but due to its smaller field detects few
groups in this mass range.

In addition to probing the X-ray properties of optically-selected groups,
the multi-wavelength coverage of this Bootes field opens the door for
additional studies.  We will be able to constrain the group velocity
dispersion and compare to X-ray mass estimators.  With the existing deep,
multi-band imaging, we will also investigate the role of environment (local
galaxy density, early-type fraction, brightest group galaxy) in determining
group X-ray properties (X-ray luminosity, gas temperature).

\section{Acknowledgements}

We thank the conference organizers for arranging an enjoyable and
stimulating meeting in a fantastic locale.  EDM acknowledges support from
NASA grant NNX08AZ64G.

\section*{References}

\re
Balogh, M.~et al.~2006, MNRAS, 366, 624

\re
Barkhouse, W.A.~et al.~2006, ApJ, 645, 955

\re
Bauer, F.E.~et al.~2002, AJ, 123, 1163

\re
Burenin, R.A.~et al.~2007, ApJS, 172, 561

\re
Donahue, M.~et al.~2001, ApJ, 552, L93


\re
Fassnacht, C.D.~et al.~2008, ApJ, 681, 1017

\re
Finoguenov, A.~et al.~2007, ApJS, 172, 182

\re
Grant, C.~et al.~2004, ApJ, 610, 686

\re
Jannuzi, B.T.~\& Dey, A.~1999, ASP~191, ed.~Weymann et al., 111

\re
Jeltema, T.~et al.~2006, ApJ, 649, 649

\re
Kenter, A.~et al.~2005, ApJS, 161, 9

\re
Lubin, L.M.~et al.~2004, ApJ, 601, L9

\re
Markevitch, M.~1998, ApJ, 504, 27

\re
Mulchaey, J.~2000, ARAA, 38, 289

\re
Murray, S.~et al.~2005, ApJS, 161, 1

\re
Osmond, J.~\& Ponman, T.~2004, MNRAS, 350, 1511

\re
Rasmussen, J.~et al.~2006, MNRAS, 373, 653

\re
Tully, R.~1987, ApJ, 321, 280

\re
Willis, J.~et al.~2005, MNRAS, 363, 675

\label{last}

\end{document}